\documentstyle[preprint,aps]{revtex}
\begin{document}

\draft

\preprint{$
\begin{array}{l}
\end{array}
$}

\title{A Model of Low-lying States in Strongly Interacting\\
Electroweak Symmetry-Breaking Sector}

\author{Tao Han$^a$\footnote{than@ucdhep.ucdavis.edu},
Zheng Huang$^b$\footnote{Present address: Department of
Physics, University of Arizona, Tucson, Arizona; 
huang@physics.arizona.edu},
and P.Q.\ Hung$^c$\footnote{hung@athena.phys.virginia.edu} }

\address{{$^a$Department of Physics, University of California,
Davis, CA 95616, USA}\\
{$^b$Theoretical Physics Group,
 Lawrence Berkeley Laboratory, Berkeley, CA 94720, USA}\\
{$^c$Department of Physics, University of Virginia,
Charlottesville, VA 22901, USA}}
\maketitle
%
%
\begin{abstract}
It is proposed that, in a strongly-interacting electroweak sector,
besides the Goldstone bosons, 
the coexistence of a scalar state ($H$) and
vector resonances such as $A_1$ [$I^G(J^P)=1^-(1^+$)],
$V$ [$1^+(1^-)$] and $\omega_H^{}$ [$0^-(1^-)$] is required 
by the proper Regge behavior of the forward scattering amplitudes. 
This is a consequence of the following well-motivated assumptions:
(a). Adler-Weisberger-type sum rules and the
superconvergence relations for scattering amplitudes
hold in this strongly interacting sector;
(b). the sum rules at $t=0$ are saturated by 
a minimal set of low-lying states with appropriate
quantum numbers.
It therefore suggests that a complete description should include 
all these resonances.
These states may lead to distinctive experimental
signatures at future colliders.
\end{abstract}


\newpage

Despite the extraordinary success of the Standard Model (SM) in
describing particle physics up to the highest energy available today,
the mechanism for the electroweak symmetry-breaking remains one of
the most prominent mysteries.
It has been argued that if there is no light Higgs boson ($H$)
with a mass less than about 1 TeV, the electroweak
symmetry-breaking sector would become strongly
interacting \cite{lqt}
and  new physics must enter.
However, it is commonly believed that a heavy Higgs boson
may be very broad due to the enhanced couplings among the $H$ and
the longitudinal vector bosons, while a $\rho$-like resonance
(denoted by $V$ henceforth)
may have a mass close to 2 TeV based on a
naive scaled-up version of QCD in various Technicolor models \cite{tc}.
Those features would make the direct experimental searches for
those heavy particles at future colliders difficult and
sophisticated acceptance cuts must be applied to enhance the
signals over  SM backgrounds \cite{search}.

In this Letter  we would like to argue that the above
conventional wisdom on the scalar and vector states
may not be  true in general.
In particular,  the separate consideration on
each individual low-lying state in a strongly-interacting
sector may  be incomplete.
The well-known example
is the strong interactions of hadrons  in  which
the requirement of good high energy behavior
of  $S$-matrix elements gives rise
to a set of Adler-Weisberger (AW)-type sum rules \cite{adler}
for forward scattering amplitudes \cite{gilman}.
To satisfy  these sum rules,  besides the Goldstone bosons
$\pi's$, the other states such as $\rho, \sigma$, $a_1$
and $\omega$ are required to coexist.
We thus propose a scenario for the
strongly-interacting electroweak sector
which contains a rich spectrum of low-lying  states,
including the Goldstone bosons ($w^\pm,z$
or  equivalently the  the longitudinal components of the weak bosons
$W_L^{}$ in high energy limit), a scalar resonance $H$,
a  $\rho$-like vector resonance $V$, and other vector
resonances such as $A_1$ and $\omega_H^{}$ \cite{omiga}
(analogous to $a_1$ and $\omega$ in low energy hadron physics).
The AW-type sum rules and the superconvergence relations
are applied to
$W_L^{}  W_L^{}$, $W_L^{} A_1$ and $W_L^{} V$ scatterings to obtain the
relations among the couplings and masses, which are fully expressed
by two parameters: a mixing angle and one of the masses.
Motivated by this rather coherent picture,
we construct an effective chiral
Lagrangian containing those resonances with  a manifest
$SU(2)_L \otimes SU(2)_R$  symmetry
(stontaneously broken down to $SU(2)_V$, the weak-isospin,
by the scalar vacuum expectation value $v$). 
The $\omega_H^{}$ interactions with
other states are separately introduced. 
As a result, many
relations  among the masses  and coupling constants
obtained by the AW-type sum rules and superconvergence relations
are also preserved in the chiral Lagrangian formalism.
The scattering amplitudes obtained from the chiral Lagrangian
can thus satisfy the sum rules as required by
a proper high energy behavior.
%

An important  consequence is  that
these low-lying states are in general quite narrow,
with widths typically of 100 GeV or less. Even  for the scalar state $H$,
its width is significantly smaller than  that in the SM due to the coexistence
of a vector state $V$. Also,  $A_1$ and $\omega_H^{}$ decay
substantially to three $W_L^{}$'s due to the conservation of
angular momentum and isospin. These features may lead to
distinctive experimental signatures at the next
generation of colliders, such as the Large Hadron Collider (LHC),
a TeV $e^+e^-$ Next Linear Collider (NLC) and a possible $e\gamma$ collider.
Of equal importance, because of the non-decoupling nature to the SM physics
at low energies, those low-lying states may manifest themselves through
virtual effects. One may therefore be able to explore the parameter space
of the model and  to test it by precision
electroweak data well before the operation of future colliders.

Before discussing our model in any detail, we recall that,
for strong interactions of hadrons, it has been long shown by
Gilman and Harari \cite{gilman} and by Weinberg \cite{sw1} that {\em all}
Adler-Weisberger-type sum rules can be satisfied with just four meson
states, the $\pi$, $\rho$, $\sigma$ and $a_1$. It was recently pointed out
by Weinberg \cite{sw2} that this is a natural result of the algebraic structure
of broken symmetry (a ``mended symmetry'') SU(2)$\otimes$SU(2), and these
four mesons should actually form a more-or-less complete representation
of the group. The mass splitting is accounted for by the non-commutativity
between the group generators and the mass matrix. Such an approach
is quite successful in relating various couplings and masses in low energy
hadron interactions.

We therefore propose that similar sum rules also apply
to the strongly-interacting
electroweak sector and thus construct a model of possible
low-lying states which mimic the algebraic structure in strong
interactions.
If we ignore the gauge interactions and the isospin breaking
due to mass splitting within fermion multiplets,
the parity ($P$) and the weak isospin ($I$) are conserved in the
electroweak symmetry-breaking  sector.
If we assign the quantum numbers $I^G(J^P)=1^-(0^-)$ to
the Goldstone bosons ($w^\pm,z$) and $0^+(0^+)$ to the Higgs boson
$H$, the quantum numbers for  the $\rho$-like resonance ${V}$
and the axial vector $A_1$ would be $1^+(1^-)$  and $1^-(1^{+})$,
respectively.

The algebraic aspects
of the symmetry precisely arise from the need for cancellation which
ensures a reasonable asymptotic behavior  of  the $S$-matrix elements
at high energies as required by Regge
theory or a renormalizable theory.
As a first test of the consistency of the theory, one may consider
the process $W_LA_1\rightarrow W_LA_1$.
The Regge theory on the high energy behavior of the forward ($t=0$)
scattering amplitude is best represented in the following
dispersion relations:
the superconvergence for $I_t=2$  in the $t$-channel \cite{gilman}
\begin{equation}
\int_0^\infty \frac{d\nu}{\nu}{\rm Im} {\cal T}^{(I_t=2)}(\nu,t=0)=0\; ,
\label{sr1}
\end{equation}
and the usual Adler-Weisberger sum
rule \cite{adler} for $I_t=1$ in the $t$-channel
\begin{equation}
\frac{2}{\pi}\int_0^\infty \frac{d\nu}{\nu^2}{\rm Im}
{\cal T}^{(I_t=1)}(\nu,t=0)=\frac{4}{\upsilon^2}\;  . \label{sr2}
\end{equation}
Here $\nu$ is the c.m. momentum squared
and $v = 246$ GeV. 
Eq.~(\ref{sr1}) is derived by assuming
 the absence of $I=2$ states (which is true in strong interactions) so that
the amplitude satisfies an unsubtracted dispersion relation in $\nu$
and its value at threshold vanishes.
The exsistence of $H$ and $V$ states is sufficient
to saturate the above sum rules.
The aforementioned scenario can be applied to $W_L V$ scattering.
One finds that the saturation of forward scattering {\it requires}
the existence  of an isosinglet vector boson, the $\omega$-like state,
to be denoted by $\omega_H^{}$ [$0^-(1^-)$].

We can derive  eight
Adler-Weisberger type sum rules for our minimal content of the low-lying
states in the strongly interacting electroweak sector:
two for $W_L A_1$ scattering  with contributions from  $H$ and $V$;
two for $W_L W_L$  scattering saturated by $H$ and $V$;
four for $W_L V$ scattering  with contributing states $W_L,~\omega_H^{}$
and $A_1$.  There are no sum rules for the scattering
of $W_L$'s on any of the isosinglet states.
The complete list of these sum rules can be found in
Refs.\cite{gilman}  and \cite{hhh}.
These sum rules yield the following relations:
\begin{eqnarray}
\label{masses}
M_V^2=M_{\omega}^2=M_H^2\tan^2\psi + M_Z^2 (1-\tan^2\psi)
 \quad  ; \quad
M_A^2=\frac{M_V^2}{\sin ^2\psi} - M_Z^2\cot^2\psi ;\\
\label{gone}
g_{\rm HWW}^2=\frac{4}{\upsilon^2}\sin^2\psi \quad ;\quad
g_{\rm \omega VW}^2=\frac{4}{\upsilon^2} \quad ;\quad
\frac{g_{\rm VWW}^2}{M_V^2}=\frac{1}{\upsilon^2}\cos^2\psi ;\\
\frac{g_{\rm AHW}^2}{M_A^2}=\frac{1}{\upsilon^2}\cos^2\psi
\quad ; \quad
g_{\rm AVW}^2 = \frac{16}{\upsilon^2} \frac{M_A^2M_V^2}{(M_A^2-
M_V^2)^2}\sin^2\psi\; ,
\label{gtwo}
\end{eqnarray}
where $\psi$ is  a free parameter, the  mixing angle between $W_L$ and
$A_1$. None of these states can be eliminated from the sum rules
without leading to apparent contradictions. For  instance, if  the $V$ state
is removed from the $W_LA_1$ sum rules by assuming the degeneracy
$M_H=M_A$ which corresponds to the ``unmixed'' case $\psi =0$,
this would then imply a grave violation of the $W_LW_L$ sum
rules \cite{gilman,sw2,hhh}.
The importance of these results is that the masses and couplings of these
low-lying states are completely parameterized by the mass of any one of
these states ($e.g.$, $M_H$) and a mixing angle $\psi$.

The canonical value of $\psi$ is $\pi/4$ in a QCD-like theory as
implied by the KSRF relation \cite{ksrf}. It then follows
$M_V\simeq M_A/\sqrt{2}$, the famous relation first derived
from the Weinberg sum rule.
Note  that in this case the scalar width is reduced
by about a half compared to that obtained from
perturbation theory in the SM.
For $\psi =\pi/4$ and some typical values of $M_H$, we summarize our
predictions in Table \ref{tab}.
In calculating the $\omega_H^{}$ width, we have assumed the $V$
exchange dominance \cite{gell-mann}.
Indeed, these states are in general quite narrow, with widths
typically of 100 GeV or less.
(Not listed in Table \ref{tab} are predictions for higher mass values,
where e.g. for $M_{\omega} \cong 1.8$ TeV, the two-body decay width
is approximately 10 \% of the three-body one.)
However, we emphasize that the deviation of $\psi$ from $\pi/4$ is
quite possible in a theory other than the QCD-like ones ({\it e.g,}
in some Extended Technicolor  models \cite{tc})
where these sum rules are expected to
hold since they are based on a more general ground.
Our parametrization allows one to determine the mixing angle
from experiments.
A preliminary exploration shows that the current precision electroweak
data seem to
favor certain region of the parameter space with moderate mixing
in our model \cite{hhh}.

In order to further study the phenomenology and to gain
insight of the underlying physics for the aforementioned
model,  we  construct an effective chiral Lagrangian
containing those low-lying states.  Guided by  the non-linear
$\sigma$-model including vector mesons in hadron strong
interactions  \cite{nonlnr}, we  parameterize the
massless Goldstone bosons $w_a$ ($a$=1,2,3)
and the vector fields by
\begin{eqnarray}
U = exp(i  \tau_a w_a/v); \; A_{L(R)}^\mu = {1 \over 2}
 (V^\mu \pm A^\mu + \omega_H^\mu),
\end{eqnarray}
%
where $V^\mu=\tau_a V^\mu_a$,  $A^\mu=\tau_aA^\mu_a$
(with  $Tr \tau_a \tau_b= 2\delta_{ab}$).
They transform linearly under  $SU(2)_L \otimes SU(2)_R$ as
\begin{eqnarray} \nonumber
&A_L^\mu  & \rightarrow  L A_L^\mu L^\dagger; \;
A_R^\mu \rightarrow  R A_R^\mu R^\dagger; \; \\
&U &  \rightarrow L U R^\dagger; \;
\omega^\mu_H  \rightarrow \omega^\mu_H; \;
H \rightarrow H.
\end{eqnarray}
Ignoring the electroweak gauge couplings,
the chiral Lagrangian realized in Goldstone mode
can then be written as
\begin{eqnarray} \nonumber
{\cal L}^{eff} = &&\frac{1} {4} v^2 \beta Tr(D^\mu U^\dagger  D_\mu U )
                          - \frac{1} {4} Tr(A_L^{\mu\nu}  A_{L\mu\nu}
                                                +  A_R^{\mu\nu}  A_{R\mu\nu}) \\  \nonumber
                   + &&\frac{1} {2} M_V^2 Tr(A_{L\mu}^2 + A_{R\mu}^2)
                   + \frac{1}{2} \partial^\mu H \partial_\mu H
                   - \frac {1} {2} M_H^2 H^2 +
\frac{1}{2} v \frac{H}{\sqrt{\beta}} Tr(D^\mu U^\dagger  D_\mu U)\\
        - &&\frac{1}{4} \lambda g^{}_V v H Tr[i(A^\mu_L U - U A^\mu_R)
          D_\mu U^\dagger + h.c.] + {\cal L}_{\omega_H},
\label{leff}
\end{eqnarray}
where
$$
D^\mu U = \partial^\mu U - ig^{}_V A^\mu_L U + ig^{}_V UA_R^\mu \ \ 
{\rm and } \ \ 
A^{\mu \nu}_{L(R)} = \partial^\mu A^\nu_{L(R)} - \partial^\nu A^\mu_{L(R)} 
- i g_V^{} [A^\mu_{L(R)}\ , A^\nu_{L(R)}] 
$$
with $g^{}_V$ the new strong coupling  constant,
which we have taken to be the same for $A_L$
and $A_R$ fields since parity is conserved in the new
strong interactions.
Parameters $\beta$ \cite{lee} and $\lambda$ are to be fixed by
proper normalization and sum rules. The 
$\omega^{}_H$ interactions with
other states can be conveniently written as \cite{nonlnr}
\begin{eqnarray}
{\cal L}_{\omega_H} =&& -5 C g_V^{} \omega^\mu_{H} \ \epsilon_{\mu \nu \alpha \beta}
Tr[L^\nu L^\alpha L^\beta]
 -\frac{15}{4} C g_V^2 \omega^{\mu \nu}_H \epsilon_{\mu \nu \alpha \beta}
Tr[V^{\alpha} (R^{\beta} - L^{\beta}) \nonumber \\
&&+ A^{\alpha} (R^{\beta} + L^{\beta})
+ V^{\alpha} A^{\beta} - \frac{1}{2} (V^{\alpha} - A^{\alpha})
U^\dagger (V^{\beta} + A^{\beta}) U].
\end{eqnarray}
Here $C$ is an arbitrary constant to be fixed by the sum rules and
$L_\mu = U^\dagger \partial_\mu U = - U^\dagger R_\mu U$. (In QCD,
$C = N_c / 240 \pi^2$.) 

Diagonalizing the quadratic crossing term in $\vec{A}_\mu$
and $\tilde{D}_\mu \vec{w}$ 
($\equiv \partial_\mu \vec{w} + g_V^{} \vec{V}_\mu
\times \vec{w}$) in Eq.~(\ref{leff}), one obtains
\begin{equation}
\beta = M_V^2/(M_V^2 - g_V^2 v^2), \ \
M_A^2 = M_V^2 + \beta g_V^2 v^2.
\end{equation}
Upon defining
\begin{equation}
\beta \equiv \frac{1}{\sin^2 \psi}, \ \
\lambda \equiv 2 (\ \frac{1}{\sin\psi} - \sin \psi), \ \
M_H^2 \equiv \beta g_V^2 v^2,
\end{equation}
the mass relations in Eq.~(\ref{masses}) and those 
for $g_{\rm HWW}^{}$ and $g_{\rm VWW}^{}$ 
in Eq.~(\ref{gone}) immediately follow.
The sum rule relation for $g_{\rm AHW}^{}$ can be easily obtained
as well\footnote{To derive the
proper Feynman rule for the AHW interaction, an integration by parts
was used to get a term like 
$\vec{A}_{\mu} \cdot \vec{\pi} \ \partial^\mu H$.}.
The second term of ${\cal L}_{\omega_H}$ 
in Eq.~(\ref{leff}) gives
the sum rule for $g_{\omega {\rm VW}}^{}$ if we choose
\begin{equation}
C \equiv \frac{1}{15 g_V^2}.
\end{equation}

It is very interesting to note that with such a simple effective
Lagrangian of Eq.~(\ref{leff}), many relations among masses and couplings
in Eqs.~(\ref{masses})-(\ref{gtwo}) obtained by strong scattering sum rules
are preserved. The only exception is
the last one in Eq.~(\ref{gtwo}) involving $g_{\rm AVW}^{}$ which will need 
some more careful treatment.
We shall address this point in a later work \cite{hhh}.


Unlike most of the previous studies \cite{search,bess},
which makes no reference to the dynamical relationship and
detail cancellation among the proposed resonances,
our scenario is based on a consistent parametrization of low-lying states
with certain internal symmetry,
obeying strong interaction sum rules as required by
a proper high energy behavior.
These low-lying narrow states could lead to  distinctive
experimental signatures at future colliders.
Even without direct couplings to fermions, the vector states $V$,
$A_1$ and $\omega_H^{}$ may be
copiously produced at the LHC and a TeV $e^+e^-$ NLC via
a mixing with the transversely polarized vector bosons $\gamma, Z$
or $W^\pm$. $A_1$ and $\omega_H^{}$ may also be produced via the fusion processes
$W_L^{} \gamma \rightarrow A_1, \omega_H^{}$, most conveniently at an $e\gamma$
collider. The dominant three-body decays of $A_1$ and $\omega_H^{}$ would provide
distinctive signature from the two-body decays of $V$.  For instance at
the LHC for $\sqrt s=14$ TeV, the production cross section for a 900 GeV $H$
is about 0.2 pb; while that for the vector states is of  order 1 pb. With the
designed annual integrated luminosity of 100 fb$^{-1}$, one would expect
$10^4 - 10^5$ 2$W_L^{}$ or 3$W_L{}$ final states. With the narrow widths for the
vector states, as well as the significantly reduced
width for $H$, the experimental identification
for these low-lying states at the LHC  should be quite promising.
Comprehensive phenomenological studies of these states
will be reported elsewhere \cite{hhh}.

To summarize, we propose that  in a strongly-interacting electroweak
sector, to satisfy the Adler-Weisberger-type sum rules
and the superconvergence relations, there exists
a rich spectrum, including the Goldstone bosons $w_a$,
a scalar $H$, a vector $V$, an axial vector $A_1$ and
an isospin-singlet vector $\omega_H^{}$, obeying certain
algebraic structure.
By applying
the AW-type sum rules, relations among masses and couplings
of those states are established and they are fully expressed
by two parameters: a mixing angle $\psi$ and one of the masses
(say, $M_H$).
It is found that in general those states are fairly narrow,
due to the necessary coexistence of the states.
Our model yields distinctive predictions and
experimental signatures.
Motivated by this model, we construct an effective chiral
Lagrangian for these states with a manifest
$SU(2)_L \otimes SU(2)_R$ symmetry.  Resultant mass and
coupling constant relations satisfy almost all the sum rules.
This effective Lagrangian can be easily applied to future
phenomenological studies. It is crucial to test these ideas
at the forthcoming LHC and/or NLC facilities and to explore the
fundamental mechanism for the electroweak symmetry-breaking.
\acknowledgements
%
T.H. and P.Q.H would like to thank the LBL theory
group for warm hospitality during the initial stage of this work.
This work was supported in part by
the U.S.\ Department of Energy
under Contracts No.\ DE-AC03-76SF00098, DE-A505-89ER40518, DE-FG03-91ER40647
and by the Natural Sciences
and Engineering Research Council of Canada.
%
%

%
%
%
\nopagebreak
\begin{table}
\caption{Predictions on the masses and widths (in units of GeV)
of $A_1$, $V$ and $\omega_H^{}$ states for the mixing angle
$\psi=\pi/4$ and  some input values of  $M_H$.}
\label{tab}
\begin{tabular}{lccc}
 $M_H$ & 800 & 1000 & 1200\\
 $\Gamma(H\rightarrow W_LW_L)$ & 120 & 238 & 416\\  \tableline
 $M_A$  & 1131 & 1414 & 1697\\
 $\Gamma(A_1\rightarrow VW_L)$  & 31.5 & 67 & 121\\
 $\Gamma(A_1\rightarrow HW_L)$  & 8.5 & 17.6 & 31.3\\ \tableline
 $M_V$  & 800 & 1000 & 1200\\
 $\Gamma(V\rightarrow W_LW_L)$ & 26.1 & 52.4 & 92\\ \tableline
 $M_\omega$ & 800 & 1000 & 1200\\
 $g_{\rm \omega VW}$ (GeV$^{-1}$) & 8.1$\times 10^{-3}$ &
8.1$\times 10^{-3}$ & 8.1$\times 10^{-3}$ \\
$\Gamma(\omega_H^{}\rightarrow Z_L W_T^3)$  & 2.2 & 2.8 & 3.3\\
$\Gamma(\omega_H ^{}\rightarrow Z_L W_L^+ W_L^-)$  & 1.5 & 4.1 & 8.9
\\
\end{tabular}
\end{table}

\end{document}